\documentstyle[aps]{revtex}
\begin{document}

\title{Quantum Hamilton-Jacobi equation }
\author{Vipul Periwal}
\address{Department of Physics,
Princeton University,
Princeton, New Jersey 08544}

\def\dd{\hbox{d}}
\def\tr{\hbox{tr}}\def\Tr{\hbox{Tr}}
\def\ee#1{{\rm e}^{{#1}}}
\def\part{\partial}
\def\Part#1{\partial_{{#1}}}
\def\Ddot{{\dd \over {\dd t}}}
\def\eps{\epsilon}
\maketitle
\begin{abstract} The nontrivial transformation  of the phase space path 
integral measure under certain  discretized analogues of canonical 
transformations is computed.
This Jacobian  is used to
derive a quantum analogue of the Hamilton-Jacobi equation  for 
the generating function of a canonical transformation that maps 
any quantum system to a system with a vanishing Hamiltonian.  A formal
perturbative solution of the quantum Hamilton-Jacobi equation is given.
\end{abstract}
A remarkable formulation of classical dynamics is provided by the 
Hamilton-Jacobi equation:
If $S(q,P,t)$ satisfies  
\begin{equation}
{{\partial S}\over{\part t}}(q,P,t) + H(q,\Part qS,t) = 0\ ,
\label{hj}
\end{equation}
where $H$ is the Hamiltonian, then the canonical transformation defined by
\begin{equation}
\Part PS = Q,\qquad \Part qS = p
\label{fix}
\end{equation}
maps the dynamical system governed by the Hamiltonian $H$ to a trivial dynamical 
system, one with vanishing Hamiltonian.  To see this, note that
\begin{equation}
p\dot q - H = \Part qS \dot q -H = \Ddot {(S-PQ)} + P\dot Q\ ,
\end{equation}
using eq.~\ref{hj}. 
Boundary terms do not affect the phase space equations of motion, so 
this mapping determines identical classical dynamics\cite{history}.  
The function $S$ is Hamilton's principal function, or action, which
acquires a  greater significance in quantum 
mechanics\cite{dirac,feynman}.

Quantum mechanically, canonical
transformations of the form considered above do not generate equivalent
quantum systems\cite{faddeev,zj,edwards}. 
 There is no natural action of the   group 
of symplectomorphisms on the quantum Hilbert space.  Alternatively,
in Feynman's formulation of quantum mechanics\cite{feynman},
the phase space path integral is not invariant under canonical 
transformations.   The non-invariance of   phase space (and  
co\"ordinate space) path integral measures has been 
the focus of a great deal of work\cite{edwards}.  In the present work, 
the general problem  of symplectic transformations will not be 
considered---I shall just consider the properties of the
phase space path integral under the discretized analogues of 
canonical transformations of a particular type.   The motivation is to 
answer  the following: Is there a 
deformation of eq.~\ref{hj} which allows a quantum mechanical map 
from an arbitrary quantum system to one with a vanishing Hamiltonian?
This question has attracted some attention in the recent 
literature\cite{ferraro,faraggi}.

After a short review  of
the path integral formulation to make the measure precise, I will
compute the   transformation of the measure under the 
transformations that keep the discretized $\int p\dd q$ term in the action 
invariant (up to total derivatives).  These 
transformations differ from canonical transformations 
due to the discretization of the phase space path 
integral, so the Jacobian for the change of variables in the path 
integral is nontrivial.  
 An important consistency check is satisfied by
the result: the change is consistent with the group property
of the  
canonical transformations considered, in the continuum limit.  
A particular application of this 
result gives  the desired  deformation of the 
Hamilton-Jacobi equation, with deformation parameter the Planck 
constant.   
From this,  the quantum
Hamilton-Jacobi equation, eq.~\ref{qhj}, is immediate.  The solution of
eq.~\ref{qhj} as a formal perturbative series takes a  
simple form, eq.~\ref{series}.

We compute $\langle q'',t''|p',t'\rangle$ as a functional integral,
choosing the momentum state to position state amplitude to obtain a
symplectically invariant form for the path integral measure. Note
$\langle p|q\rangle = (2\pi)^{{-d/2}}\exp(-ipq),$
and if $H$ is ordered so that all momentum operators appear on the 
left, $\langle p|H|q\rangle = (2\pi)^{{-d/2}}\exp(-ipq)H(q,p).$
Assume that the Hamiltonian is time-independent for notational 
simplicity, since
the generalization to arbitrary Hamiltonians is trivial.
Since 
\begin{equation}
\langle q'',t''|p',t'\rangle = \lim_{N\uparrow\infty}\langle 
q''|(1-i\eps H)^{N}|p'\rangle\ ,
\end{equation}
with $\eps\equiv (t''-t')/N,$ using
$1=\int \dd p \dd q |p\rangle\langle q| (2\pi)^{{-d/2}}\exp(-ipq) $
between every factor of $(1-i\eps H),$
we find 
\begin{equation}
 \langle q'',t''|p',t'\rangle = {1\over \sqrt{2\pi}}\lim_{N\uparrow\infty} 
 \int \prod_{{i=1}}^{N} {{\dd p_{i} \dd q_{i} }\over (2\pi)^{d}}
 \ee {iA_{N}}\ee {ip_{0}q_{1}}\ ,
\label{precise}
\end{equation}
where $A_{N}\equiv \sum_{i=1}^{{N}}\left[ p_{i}(q_{i+1}-q_{i}) -\eps 
H(p_{i},q_{i})\right].$  Here, $q_{N+1}\equiv q''$ 
and $p_{0}=p',$ and $q_{1}$ and $p_{N}$ are integrated over.
In the continuum 
limit, $A_{N}\rightarrow A_{\infty}\equiv \int \dd t \left[p\dot q - 
H\right],$ and the measure can be described   heuristically as 
an integration over all
phase space paths satisfying $q(t'')=q'', p(t')=p,$ with
$p(t'')$ and $q(t')$ 
integrated over.  For the pitfalls in such continuum descriptions, see 
\cite{faddeev,zj,edwards}.

Eq.~\ref{precise} can now be used to consider the properties of 
the phase space path integral under canonical transformations. 
The measure $\prod \dd p_{i} \dd q_{i}$
is clearly invariant under arbitrary $i$--dependent canonical 
transformations as a straightforward
mathematical fact.  However, $A_{N}$ is not invariant under 
such transformations.  The point of the following exercise is to 
find a transformation of integration variables $(p_{i},q_{i}) 
\rightarrow (P_{i},Q_{i})$  that changes the $p\dd q$
term in  $A_{N} $ in a simple way,  and then to 
compute the Jacobian for this transformation.   

Consider defining functions $Q(q,p),P(q,p)$ implicitly by means of 
the following definitions, for arbitrary functions $S_{i}(P,q):$
\begin{equation}
p_{i}(q_{i+1}-q_{i}) \equiv S_{i}\left(q_{i+1},P_{i}) - 
S_{i}(q_{i},P_{i}\right)\ ,\qquad Q_{i}(P_{i}-P_{i-1}) \equiv 
S_{i-1}\left(q_{i},P_{i}) - 
S_{i-1}(q_{i},P_{i-1}\right)\ .
\label{PQS}
\end{equation}
Now observe that 
\begin{equation}
p_{i}(q_{i+1}-q_{i})+Q_{i}(P_{i}-P_{i-1}) = \left[S_{i} (q_{i+1},P_{i}) -
S_{i-1}(q_{i},P_{i-1} )\right] -\left[S_{i} (q_{i},P_{i}) -
S_{i-1}(q_{i},P_{i} )\right]\ ,
\label{observ}
\end{equation}
with the first term in $[\dots]$ a telescoping series when summed 
over $i.$  Note that eq.~\ref{observ} has no dependence on $H.$
Thus one finds
\begin{equation}
A_{N} = \sum_{i=1}^{{N}}\left[ -Q_{i}(P_{i }-P_{i-1}) -\eps 
H(p_{i},q_{i})-\left\{S_{i} (q_{i},P_{i}) -
S_{i-1}(q_{i},P_{i} )\right\}\right]  + \hbox{boundary terms}.
\label{almost}
\end{equation}
Comparing eq.~\ref{almost} with eq.~\ref{hj},  this is
the form expected if time is discretized.  I must now compute the
effect of the substitutions in eq.~\ref{PQS}  on the measure.

Keeping $P_{i-1},q_{i+1}$ fixed, I find that 
\begin{equation}
\dd p_{i}\dd q_{i} = (q_{i+1}-q_{i})^{{-1} }\Part {P_{i}}\left[
S_{i}(q_{i+1},P_{i})-S_{i}(q_{i},P_{i})\right] \dd P_{i}\dd q_{i}\ ,
\label{dpdq}
\end{equation}
whereas
\begin{equation}
\dd P_{i}\dd Q_{i} = (P_{i}-P_{i-1})^{{-1} }\Part {q_{i}}\left[
S_{i-1}(q_{i},P_{i})-S_{i-1}(q_{i},P_{i-1})\right] \dd P_{i}\dd q_{i}\ .
\label{dPdQ}
\end{equation}
The Jacobian for the change of variables $(p,q)_{i} \rightarrow (P,Q)_{i}$
is therefore non-trivial. It is not possible to proceed further
without some knowledge of the relation between the canonical variables with
subscripts $i$ and the variables with subscripts $i\pm 1,$ in other 
words, without some restriction on the sequences $q_{i}$ and $P_{i}$ 
as $N\uparrow \infty.$  I will come back to these restrictions 
momentarily.

At a formal level, {\it assuming that $P_{i-1}-P_{i}$ and $q_{i+1}-q_{i}$ 
are
small as $N\uparrow \infty$}, it follows from eq.'s~\ref{dpdq},\ref{dPdQ} that
\begin{eqnarray}
&&\dd p_{i}\dd q_{i} =  \left[\Part 
{P_{i}}\Part {q_{i}} S_{i}(q_{i},P_{i}) + {1\over 2} (q_{i+1}-q_{i}) 
\Part 
{P_{i}} \part_{q_{i}}^{2} S_{i}(q_{i},P_{i}) +\dots \right] \dd 
P_{i}\dd q_{i}
\nonumber\\
&&\dd P_{i}\dd Q_{i} =  \left[\Part 
{P_{i}}\Part {q_{i}} S_{i-1}(q_{i},P_{i}) - {1\over 2} (P_{i}-P_{i-1}) 
 \part_{P_{i}}^{2} \Part {q_{i}} S_{i}(q_{i},P_{i}) +\dots \right] \dd 
P_{i}\dd q_{i} \ .
\label{differ}
\end{eqnarray}
We can also derive the analogue of eq.~\ref{differ} for $\dd 
q_{i+1}\dd p_{i}:$  
\begin{eqnarray}
&&\dd q_{i+1}\dd p_{i} =  \left[\Part 
{P_{i}}\Part {q_{i+1}} S_{i}(q_{i+1},P_{i}) - {1\over 2} (q_{i+1}-q_{i}) 
\Part 
{P_{i}} \part_{ q_{i+1}}^{2} S_{i}(q_{i+1},P_{i}) +\dots \right] \dd 
q_{i+1}\dd 
P_{i}
\nonumber\\
&&\dd Q_{i+1}\dd P_{i} =  \left[\Part 
{P_{i}}\Part {q_{i+1}} S_{i }(q_{i+1},P_{i}) + {1\over 2} (P_{i+1}-P_{i }) 
 \part_{P_{i}}^{2} \Part  {q_{i+1}} S_{i}(q_{i+1},P_{i}) +\dots \right]
\dd q_{i+1} \dd P_{i} \ .
\label{differtwo}
\end{eqnarray}
Eq.'s~\ref{differ},\ref{differtwo} determine Jacobians that differ by
the sign of the total time derivative contribution, indicating that 
this is a non-universal artifact of the discretization.  Such 
contributions are, of course, to be expected, since the relation of
the index $i$ to the continuum time variable $t$ for $q,P,S$ need not
be the same.  We use the ultralocality of the phase space measure to
eliminate this total derivative contribution by averaging the 
Jacobians determined by 
eq.'s~\ref{differ},\ref{differtwo}---heuristically, one can interpret
this as setting the time associated with $P_{i}$ midway between
$q_{i}$ and $q_{{i+1}}.$
So, finally, assuming that $S_{i}$ is chosen to become a differentiable function 
of $t$
as $N\uparrow\infty,$ we find 
\begin{equation} 
\lim_{N\uparrow\infty}\prod \dd p_{i}\dd q_{i}=  \lim_{N\uparrow\infty}
\prod \dd P_{i}\dd Q_{i} \cdot \exp\left({1\over 2}\int \dd t \part_{t} 
\ln\det \Part P\Part qS(q,P,t)\right) \  ,
\label{ohboy}
\end{equation}
where we use $ {\dd{}/{\dd t}} =\Part t + \dot q\Part q + \dot P\Part 
P.$  

Eq.~\ref{ohboy} has {\it exactly} 
the form that one expects, {\it in the continuum limit},
since successive canonical transformations
obey a group law that is consistent with the $\part_{t}\ln \det \Part P
 \Part q
S$ form of the Jacobian.  This is an important consistency check on
the calculation.  In hindsight, therefore, one merely needed to fix the
coefficient in front of this term.  
 
 \def\tt{(t-t')}
 We can check this Jacobian by 
performing an explicit calculation in  { any} quantum mechanics 
problem,
since the measure's transformation properties are universal, {\it 
i.e.},  independent of the Hamiltonian.
A simple choice of Hamiltonian   is $H={1\over 2} 
(p^{2}+q^{2}),$ the harmonic oscillator.
 In this case, one knows\cite{feynman} that
\begin{equation}
\langle q'',t''| p',t'\rangle = {1\over \sqrt{ 2\pi\cos\tt}}
\exp\left(-{i\over 2}\tan\tt\left[(p'^{2}+q''^{2}) 
-2p'q''\csc\tt\right]\right)
\label{fh}
\end{equation}
Choose $S(q,P,t)\equiv qP \sec(t-t') -(q^{2}+P^{2})\tan\tt/2.$  
This choice of $S$ amounts to $P = p \cos\tt + q\sin\tt,$ 
with $Q = q\cos\tt - p\sin\tt,$ and  satisfies the classical
Hamilton-Jacobi equation, eq.~\ref{hj}.   According to the 
calculations above (eq.'s~\ref{precise},\ref{almost}), 
performing some trivial integrations  
the transition amplitude should equal
\begin{equation}
 \langle q'',t''| p',t'\rangle = {1\over \sqrt{2\pi}}
 \int  {{\dd P_{N}\dd Q_{1}}\over 
 2\pi} \ee{iS(t'')} \ee{iQ_{1}(p'-P_{N})} \ee{ {1\over 2} \ln\sec\tt}\ .
 \label{trans}
 \end{equation}
 Comparing this form to eq.~\ref{fh}, we find exact agreement.
 
Eq.~\ref{ohboy} implies that 
 under the  transformation defined by 
eq.~\ref{PQS}, 
\begin{equation}
\int \dd t\left[ p\dot q - H(p,q)\right] \rightarrow \int 
\dd t\left[ p(P,Q) \dot q(P,Q) - 
H(p(P,Q),q(P,Q)) -   {i\over 2} \Part t{\ln\det \Part P\Part q 
S}\right] \ .
\end{equation}
Thus, using eq.~\ref{almost} and restoring $\hbar$, if $S$ satisfies 
\begin{equation}
\Part t\left(S+{i\over 2}\hbar {\ln\det \Part P\Part q 
S}\right) + H(q,\Part qS) = 0 \ ,
\label{qhj}
\end{equation}
eq.~\ref{PQS} will map the quantum system to a quantum system with 
a vanishing Hamiltonian.  The telescoping terms in eq.~\ref{observ}
give rise to boundary terms in the path integral of
 $\exp(iS(P(t''),q(t''),t''))$ and $\exp(-iS(P(t' ),q(t' ),t' 
 )+ip(t')q(t')).$   

What are the conditions for the validity of the formal manipulations 
that lead from eq.'s~\ref{dpdq},\ref{dPdQ} to eq.~\ref{ohboy}?  The measure 
on phase space with the 
Hamiltonian $H$ must be concentrated on paths such that $q_{{i+1}}-q_{i}$
tends to zero with $\epsilon,$ and similarly for $P_{i}-P_{i-1} $ with 
the measure determined by the transformed Hamiltonian. 
This is true with quite mild restrictions\cite{zj} on $H(p,q)$ for $q,$ 
and similar restrictions on $H'(P,Q)\equiv 
H(p(P,Q),q(P,Q)) + \Part t[ S + i/2 
\ln\det\Part P\Part q S]$ for $P.$ 
The smoothness of $P$ paths is trivially 
true  after the change of variables if $S$ 
satisfies eq.~\ref{qhj},  since the action is just
$-\int \dd t Q\dot P.$   In this context, it should be noted
that the form of the transformed Hamiltonian, $H',$  is {\it only} valid 
in the $\epsilon \downarrow 0$ limit---for finite $\epsilon,$ one must
work with the discrete forms for all quantities, including the 
substitutions for $p_{i},q_{i}$ in the Hamiltonian.  It is difficult to
make general statements about the discretized theory, as is well-known. 
 The applications of eq.~\ref{qhj} to 
field-theoretic problems may be more interesting, for ordering 
difficulties in field theory are usually absorbed into renormalization
constants\cite{zj}.   
  
Eq.~\ref{qhj} may appear to be a simple 
deformation of eq.~\ref{hj}, but in fact it is not.  According to 
Jacobi's theorem\cite{history}, finding a sufficient number of 
solutions of eq.~\ref{hj} allows one to solve the dynamics of the
system---the key point is that the variables $P$ are integration 
constants for these solutions, an interpretation possible since they 
do not appear in eq.~\ref{hj} explicitly.  This interpretation is
not possible for eq.~\ref{qhj}, so {\it a priori} one has to find appropriate
choices of $P$ before one can even attempt to solve this equation, 
unless one treats $\hbar $ as a perturbation parameter.
Since such a perturbative solution is not a good approximation in
general, 
one may be led 
to conclude that eq.~\ref{qhj} is of  less practical value in 
quantum mechanics than eq.~\ref{hj} is in classical mechanics.   Nevertheless, 
eq.~\ref{qhj} is  simple, and of   conceptual value in
understanding the classical limit of quantum mechanics.  A formal 
solution to eq.~\ref{qhj} can be found as follows: Let $S\equiv 
S_{0}+\hbar S_{1}+\hbar^{2} S_{2} +\dots.$  Then
\begin{eqnarray}
\nonumber 
&&\part_{t} S_{0}(q,P,t)+ H(q,p=\Part q S_{0},t)  =0,  \\
\nonumber
&&\part_{t} S_{1}(q,P,t)+ \Part pH(q,p=\Part q S_{0},t)\part_{q} 
S_{1}(q,P,t)  = -{i\over 2}\tr\left[(\Part P\Part q S_{0})^{{-1}}
\part_{t} \Part P\Part q S_{0}\right],   \\ 
\nonumber
&&\part_{t} S_{2}(q,P,t)+ \Part pH(q,p=\Part q S_{0},t)\part_{q} 
S_{2}(q,P,t)  = -{i\over 2}\part_{t}\tr\left[(\Part P\Part q S_{0})^{{-1}}
\part_{t} \Part P\Part q S_{1}\right], \\ 
\nonumber
&&\part_{t} S_{3}(q,P,t)+ \Part pH(q,p=\Part q S_{0},t)\part_{q} 
S_{3}(q,P,t)  = -{i\over 2}\part_{t}\tr\left[(\Part P\Part q S_{0})^{{-1}}
\part_{t} \Part P\Part q S_{2} - {1\over 2} \left(
(\Part P\Part q S_{0})^{{-1}}
 \Part P\Part q S_{1}\right)^{2} \right],   \dots \\  
\label{series}
\end{eqnarray}
The solution to this set of equations is obtained by the method of 
characteristic projections.  Let $S_{0}$ be a complete integral of
eq.~\ref{hj}, which of course coincides with the first equation in 
eq.~\ref{series}, and  $q(t)$   a solution of 
$\dot q = \part_{p}H(q(t),p=\Part qS_{0},t),$ which is just one of the
classical equations of motion.  Then $ S_{1}(q(t),P,t)$ is a solution 
of 
\begin{equation}
{\dd \over {\dd t}} S_{1} = -{i\over 2}\tr\left[(\Part P\Part q S_{0})^{{-1}}
\part_{t} \Part P\Part q S_{0}\right](q(t),P,t), 
\label{character}
\end{equation}
with analogous equations for $S_{i},i>1.$  We see, therefore, that the
integral surfaces, indexed by $P,$ of eq.~\ref{series}, depend on
the  behaviour of integral surfaces {\it as functions of } 
 $P.$ Thus, the perturbative solution of
eq.~\ref{qhj} incorporates information about quantum fluctuations by
its dependence on the complete integral of eq.~\ref{hj} at 
neighbouring values of $P.$ 
 
It would be interesting to see if
exactly solvable quantum mechanics problems can be interpreted as 
explicit solutions of eq.~\ref{qhj}.  
Eq.~\ref{ohboy} shows, further,
that the transformation to classical action-angle variables leaves
behind a non-trivial Hamiltonian, 
${i\over 2}\hbar\Part t \ln\det  \Part P\Part q S ,$
which takes into account   quantum fluctuations.  
Classical canonical transformations that solve eq.~\ref{hj}, 
and satisfy
$\Part q\det\Part P\Part qS = \Part P\det\Part P\Part qS = 0,$
will also solve the quantum dynamics, with the anomalous term serving
as a computation of the fluctuation determinant about classical 
solutions, as in the harmonic oscillator considered above.

The formulation 
considered above for canonical transformations may be  too limited.
The variables $P$
have a fundamentally different r\^ole to play in eq.~\ref{qhj} as 
compared to eq.~\ref{hj}, and it may be  natural to look for
solutions in which $P,Q$ describe  a non-commutative symplectic 
manifold.  This is suggested by the fact that the   
quantum energy spectrum could have 
discrete and/or continuous components, and   such a space cannot 
always be
described as a commuting symplectic manifold\cite{conehead}.
In such a case the form of the anomaly will be  
different. 
It would be fascinating if quantum mechanics on a 
commuting phase space could be mapped to a vanishing Hamiltonian on a
(possibly) non-commuting phase space.

To conclude, I  mention that 
two recent works\cite{ferraro,faraggi} have addressed  related issues.
In \cite{ferraro}, it is claimed that the complete solution of the
classical Hamilton-Jacobi equation, eq.~\ref{hj}, determines the 
quantum mechanical amplitude by means of a single momentum integration instead 
of a path integral.  While the path integration of the trivial quantum 
mechanics with vanishing Hamiltonian indeed reduces to (a variant of) a
phase space 
integration as 
mentioned above  (and explicitly found in the case of the harmonic 
oscillator,  eq.~\ref{trans}),    
eq.~\ref{qhj} is distinct from the classical equation, 
so it appears to contradict \cite{ferraro}.     
\cite{faraggi} postulates a diffeomorphic covariance principle, based partly on 
an SL(2,C) algebraic symmetry of a Legendre transform,  and 
finds a modification of the 
classical Hamilton-Jacobi equation that has appropriate covariance 
properties for the postulated   equivalence.  Their 
function $S$ satisfies an equation quite different from eq.~\ref{qhj},
and it is argued that $S$ is related to solutions of the Schr\"odinger equation.
Functional integrals of any sort do not appear in \cite{faraggi}, and
there is no relation to the present result, eq.~\ref{ohboy}.

I am grateful to   S. Treiman and A. Anderson for  helpful 
conversations, and I. Klebanov and W. Taylor for comments. 
This work was supported in part by NSF grant PHY96-00258.

\end{document}